# Very Low Frequency QPO in 4U1700-37

J. F. Dolan

Department of Astronomy

San Diego State University

San Diego, CA  92182-1221  U.S.A.

E-mail: tejfd@sciences.sdsu.edu

## ABSTRACT

Very low frequency ($\nu \sim$ 1.9 mHz) quasi-periodic oscillations (QPO) have been detected in 2 - 60 keV X-rays from 4U1700-37 using auto-correlation function analysis of RXTE archival data. No pulsar signal was detected in the data. Less than 0.5% of the flux from the source (3σ upper limit) is coherently pulsed with a period between 250 ms and 40 s. This upper limit corresponds to a pulsed flux <0.02 mJy. The existence of QPO in 4U1700-37 and the lack of a detected pulsar indicates that the X-ray source is more likely to be a black hole than a neutron star, consistent with recent mass estimates.

## 1. Introduction

The X-ray source 4U1700-37 is the secondary in an eclipsing binary system with an O6f primary, HD153919 (cf. Dolan et al. 1980).  Superior conjunction of the X-ray source occurs at epoch HJD 2,446,161.3409(30) + 3.417233(50) φ (Haberl, White & Kallman 1989) for integer values of orbital phase φ.  The X-ray eclipse extends from $0.93 \leq \varphi \leq 0.07$ (van Paradijs, Hammmerschlag-Hensberge & Zuiderwijk 1978).  The orbit of 4U1700-37 has semi-major axis $a_X = 31$ (+ 3, -2) $R_O$ (where $R_O$ is the solar radius) = 72 (+ 7, -5) ls, eccentricity $e_X \sim 0$ (Rubin et al. 1996) and inclination to the line of sight i= 85° ± 3° (Dolan & Tapia 1988).

4U1700-37 is not known to exhibit any periodic (pulsar) signal in its X-ray emission, but was previously considered to be a spherically accreting neutron star rather than a black hole (Dolan et al. 1980; Reynolds et al. 1999; Boroson et al. 2003; but cf. Brown et al. 1996).  Recent mass estimates of $M_X = 2.44 \pm 0.27$ $M_O$ (Clark et al. 2002; Rude et al. 2010) are significantly above the usually accepted mass limit for neutron stars, however.  We therefore analyzed 48 hours of archival RXTE photometry of the object to obtain a better upper limit on any periodic signal in the X-ray region.  We expect no periodic signal if 4U1700-37 is a black hole, as suggested by the latest mass estimates.

## 2. Observations and Data Analysis

We analyzed 48 hours of archival RXTE satellite data from 4 different epochs: 6.3 h from 1996 August 9, and September 9 and 13; 9.9 h from 1999 April 17 – 30 and May 3, 11, and 28; 22.3 h from 2000 August 20 - 23; and 10.1 h from 2003 August 30 and September 8.  Individual continuous observations ranged from 0.5 h to 4 h in length.  The Xe proportional counter array (PCA) data selected had 125 ms time resolution and a nominal 2 - 60 keV bandpass.  Detecting areas up to 6000 $cm^2$ provided counting rates between 150 – 4000 counts $s^{-1}$ from the source.  4U1700-37 appears variable on all time scales from 125 ms to tens of minutes in this X-ray bandpass (Fig. 1).

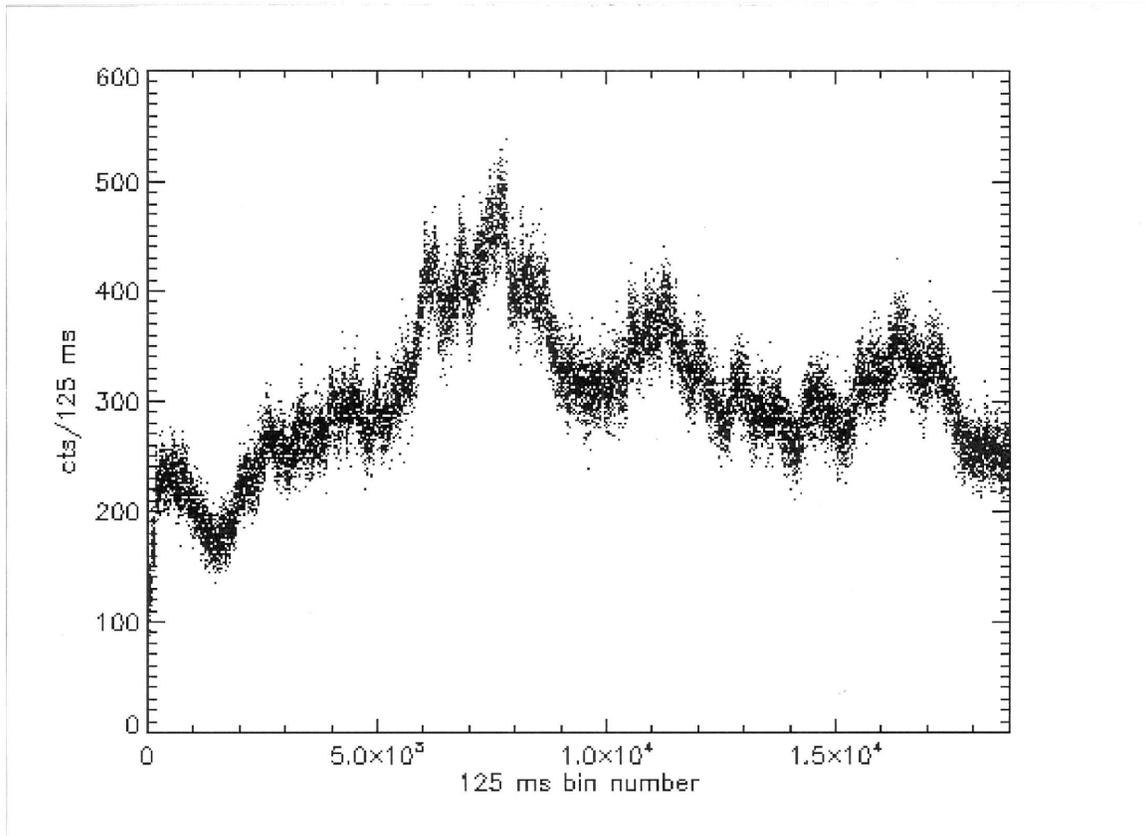

Figure 1.  The nominally 2 - 60 keV counting rate from 4U1700-37 on 1996 August  9 from 1422 to 1530 UT (J 2 450 305.09889 - .14630).  Each point represents  the number of counts detected by the PCA during a 125 ms interval.  The number of 125 ms intervals since the starting time of the 2,350 s long observation is given on the X-axis.

We searched for periodic signals using both power spectral function (PSF) and auto-correlation function (ACF) analysis. The PSF is most sensitive to signals with a sinusoidal pulse profile, while the ACF is most sensitive to delta-function signals; most physical pulse profiles lie between these two extremes. Any finite time series will have some frequency or period which contains more power in it than any other. The question is whether that power is significantly greater than that expected from a random number series having the same mean and standard deviation as the data set analyzed. We used the test of significance given by Fisher (1929; cf. also Shimshoni 1971) to determine the significance of the strongest signals in any PSF. The standard $\sigma = 1/\sqrt{N}$ criterion (Chatsworth 2004), where N is the number of data points in the series, was used to determine the significance of the strongest signals in the ACF.

We determined the sensitivity of the analysis to periodic signals by injecting synthetic signals with a period of 2 s – one a pure sinusoid, the other a delta function – into a 1996 August data set and a 2000 August data set. The mean count rate of this artificial signal was scaled by the significance of its detection to determine the $3\sigma$ upper limit on the intensity of any periodic signal in the data. Further details of these analysis procedures can be found in Percival et al. (1995).

## 3. Results

### 3.1 Periodic signals

No pulsar was detected in any data set. We found that < 0.5% of the 2 – 60 keV flux (3σ UL) is pulsed with a period between 250 ms and 40 s. This upper limit corresponds to a pulsed flux <0.02 mJy. No X-ray pulsar has ever been detected in 4U1700-37.

### 3.2 Quasi-periodic oscillations

The ACF of the data in the 2 – 60 keV bandpass taken on 2000 August 20 is shown in Fig. 2 as a function of 125 ms lag times. The two prominent local maxima, at ~4200 and ~8400 (= 2 x 4200) lags, indicate the presence of power in the data with a period near 525 s. When the data is folded modulo 525 s, a quasi-sinusoidal pulse profile emerges with a peak-to-peak amplitude ~30% of the mean count rate (Fig. 3). As indicated by the width of the local maxima in the ACF ($\Delta p/p \sim 0.25$, where $\Delta p$ is the FWHM of the ACF peak above zero), essentially the same pulse profile appears when the data is folded modulo periods between 450 s and 600 s.

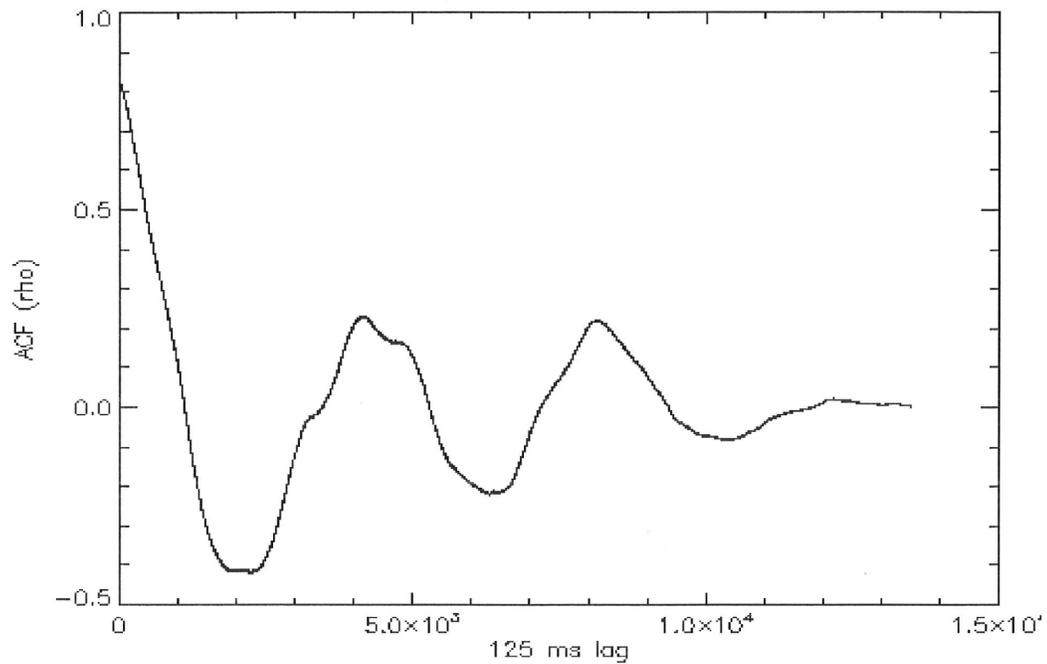

Figure 2. The ACF of the 2 - 60 keV flux from 4U1700-37 on 2000 August 20 observed between 1937 and 2006 UT (JD 2 451 777.31759 - .33727). The local maxima at ~4200 and ~8400 (and possibly ~12,600) lags correspond to power at a period of ~525 s.

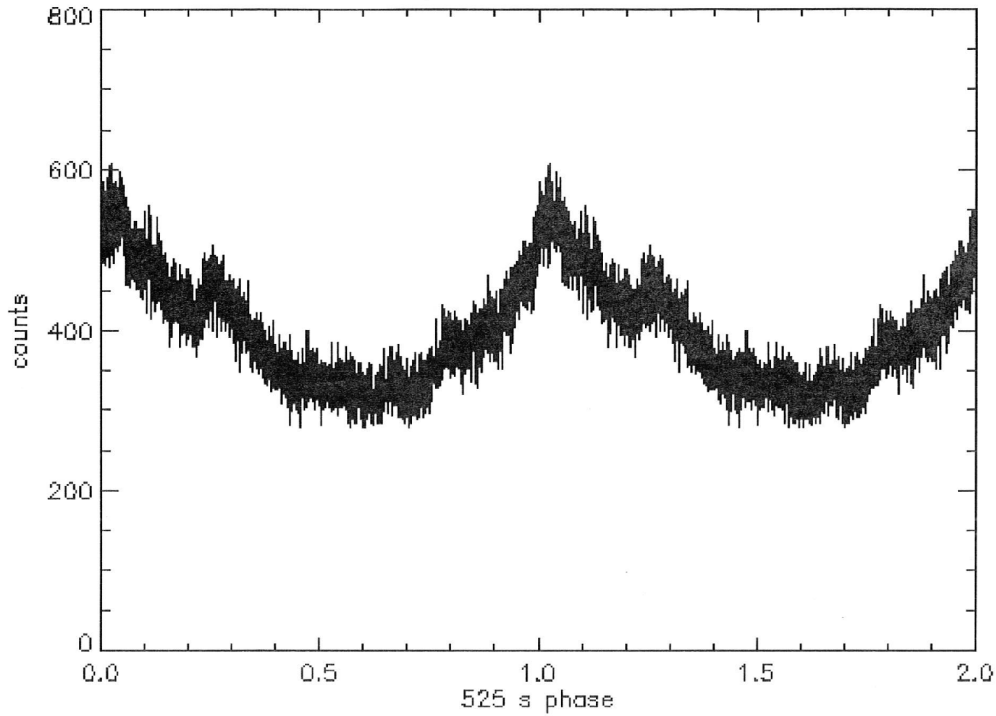

Figure 3. The data with the ACF displayed in Fig. 2 folded modulo 525 s. The arbitrary phase is defined to be zero at the start of the observation. The pulse profile is repeated twice for clarity. The secondary peak around phase 0.3 is a common feature of most QPO pulse profiles from 4U1700-37 and is apparently real.

The characteristics of this pulsation show it to be quasi-periodic oscillation (QPO). A significant peak near this period is not always present in the ACF of 4U1700-37 data, being absent in ~15% of the epochs surveyed. When a significant period is present in the ACF, folding the counts detected modulo that period gives a pulse profile with a similar shape at most epochs, but the phase of the peak of the profile is not maintained from day to day.  Coherence between pulse profile peaks is maintained for 3 – 4 hours at most, or ≤30 pulse periods.  The coherence parameter (or quality factor) of QPO is defined as $Q = \nu/\Delta\nu$ where $\Delta\nu$ is the FWHM of the QPO peak in the PSF function (Nowak 2000).  Because $|\nu/\Delta\nu| = |p/\Delta p|$, Q ~4 for the pulsation we detect, a value typical of QPO (Remillard & McClintock 2006).  QPO is clearly present in the X-ray emission of 4U1700-37.

Boroson et al. (2003) suspect that a feature below 10 mHz they detect in the PSF of a 12h uninterrupted observation of 4U1700-37 "may represent QPO with a centroid frequency of 6.5 mHz", but state that it is only "detected with marginal significance".  The authors estimate a fractional rms power of 1.4% in this feature.  Boroson et al. also note "several (other) peaks in the $10^{-3}$ to $10^{-2}$ Hz range that may be QPOs seen at low significance".  They give no numerical estimate of the statistical significance of any features.  (The reduced $\chi^2$ estimate they give refers to the acceptability of the mathematical form they assume for the PSF as a whole, not the significance of any individual peak in the PSF.)  We are unable to determine whether their results are related to the 1.9 mHz QPO we detect at many different epochs with >3σ significance.

## 4. Discussion

The question arises as to why QPO of this magnitude has not been detected previously in 4U1700-37. We believe the reason is the almost exclusive use of the PSF to analyze its X-ray photometry. The data shown in Fig. 1, for example, has QPO present with a peak-to-peak amplitude ~20% of the mean 2 - 60 keV flux and a pulse profile similar to that in Fig. 3. The frequency of this QPO lies at ~1.9 mHz, however, and is masked by the stochastic noise present below 100 mHz in the PSF of the data set (Fig. 4). An ACF analysis is more effective than a PSF analysis in detecting low-frequency (<100 mHz, i.e.,p >10 s) QPO in X-ray photometry. Shrader et al. (2010) arrive at the same conclusion in discussing their detection of QPO in three other sources.

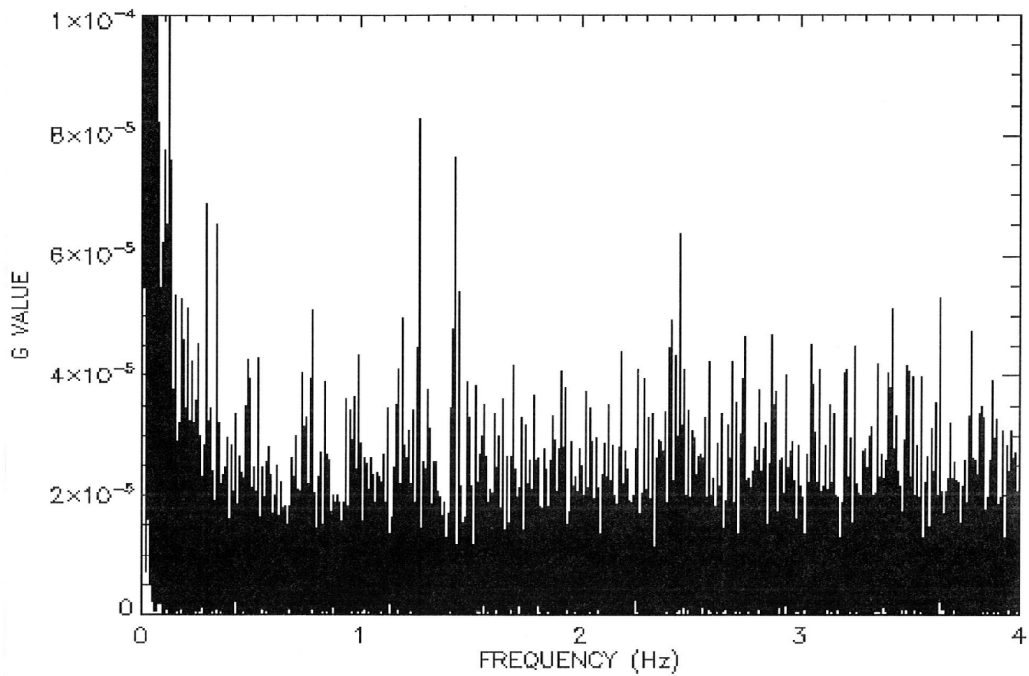

Figure 4. The PSF of the data shown in Fig. 1. The frequency resolution of the 2,350 s data set is 0.4 mHz; the Nyquist frequency is 0.85 mHz. The Fisher (1929) g value gives the significance of the power at any frequency. The 525 s period QPO peaks at ~1.9 mHz. None of the peaks above 100 mHz reach the 3σ level of significance.

1.9 mHz is the lowest frequency QPO reported from an X-ray binary. Remillard and McClintock (2006) classify low-frequency QPO as being in the range 0.1 – 30 Hz, but this classification is primarily to distinguish it from high-frequency QPO (40 - 450 Hz). QPO has been observed in several sources at frequencies significantly below 0.1 Hz (Table 1). We feel it appropriate to denote QPO with $\nu < 0.1$ Hz as very low frequency (VLF) QPO.

___

### Table 1

### X-Ray Binaries Exhibiting VLF QPO

| Source | $\nu_{QPO}$ (Hz) | Reference |
|---|---|---|
| Cygnus X-1 | 0.04 – 0.07 | Vikhlinin et al. 1994 |
| GRO J1719-24 | 0.04 – 0.3 | van der Hooft et al. 1996 |
| GRS 1915+105 | 0.067 – 1.8 | Morgan et al. 1997 |
| XTE J1118+480 | 0.07 – 0.15 | Wood et al. 2000; Revnivtsev et al. 2000 |
| 4U1700-37 | 0.0019 | this paper |

___

A circular Keplerian orbit of period 525 s around a 2.44 $M_O$ central mass object has radius a ~ $10^4$ $R_S$, the Schwarzschild radius of a black hole of that mass. VLF QPO does not appear to be associated with the inner accretion disk around a black hole (R < 20 $R_O$), where the X-ray emission is expected to arise. Remillard and McClintock (2006) have pointed out that even low-frequency QPO has frequencies much lower than the Keplerian orbital frequencies of the inner accretion disk. The physical process associated with the origin of both low-frequency and VLF QPO still awaits a theoretical explanation.

Fourteen of the 18 X-ray binaries exhibiting low-frequency QPO listed by McClintock and Remillard (2006) have been classified by them as black hole binaries, including three of the other four VLF QPO sources in Table 1. GRO J1719-24, the remaining source in Table 1, is classified as a black hole candidate by van der Hooft et al. (1996). It is reasonable to conclude that the VLF QPO detected in 4U1700-37, together with the absence of a pulsar in the system, is consistent with its identification as a black hole as indicated by the most recent spectroscopic mass estimates.

**Acknowledgements.** We thank Blaine Gilbreath for his assistance in searching the RXTE data for a pulsar signal. We acknowledge support from the Research Experience for Undergraduates Program at San Diego State University.